\documentclass[pra, aps, english, twocolumn, hyperref, superscriptaddress]{revtex4-1}

\usepackage{graphicx}

\usepackage{amssymb, amsmath, color, rotating}
\newcommand{\tb}{{\tt t}}

\begin{document}

\title{Dynamics of correlations in shallow optical lattices}

\author{Stefan S. Natu}
\email{snatu@umd.edu}
\affiliation{Laboratory of Atomic and Solid State Physics, Cornell University, Ithaca, New York 14853, USA.}
\affiliation{Condensed Matter Theory Center and Joint Quantum Institute, Department of Physics, University of Maryland, College Park, Maryland 20742-4111 USA}
\author{Erich J. Mueller}
\affiliation{Laboratory of Atomic and Solid State Physics, Cornell University, Ithaca, New York 14853, USA.}

\begin{abstract}
We explore the time-evolution of correlations in a 
homogeneous gas of lattice bosons with filling factor $n_{0}$, following a sudden reduction in the lattice depth to a regime where the interactions are weak. In the limit of vanishing interactions, we find a simple closed form expression for the static structure factor.  The corresponding real-space density correlation function shows multiple spatial oscillations which disperse linearly in time.  By perturbatively including the effect of interactions, we study the evolution of boson quasi-momentum distribution following the quench. In $1$D, the quasi-momentum distribution develops peaks at finite momentum which disperse towards $q = \pm\pi/2$. In $2$D, the momentum occupation rapidly approaches a thermal-like distribution. Quasi-long range order is never found at finite time. Our studies provide insight into the dynamics of isolated quantum systems.   
\end{abstract}
\maketitle

\section{Introduction} 
While many phenomena in nature can be qualitatively understood by simple ``mean-field"-type theories, these approaches 
inevitably capture only a subset of the interesting physics.  For example, the Mott insulating state of lattice bosons is not inert, rather there is a gas of particle-hole pairs which gives rise to a finite correlation length. Similarly, a superfluid is not fully characterized by a coherent state.  Recent experimental advances in ultra-cold gases have given us new tools for studying these fluctuations \cite{otherbragg, miyakebragg, greinerband, demarcoband, greinerdyn, blochquench, blochnoise, chengstruc, altmannoise, schneidertransport}.  A particularly promising technique (largely unique to cold atoms) is to rapidly change the Hamiltonian parameters such as hopping rate $J$ and interaction strength $U$.  The evolution following such a quench gives many insights into the single and many-particle properties of the system  ---  the spectrum of excitations \cite{chengsakh, natubogquench},  the manner in which correlations develop \cite{blochlightcone, cardy, liebrobinson}, and the role of quantum coherence \cite{greinercollapse, gerbier, orzelosc, altman}. Here we calculate how various correlation functions evolve after an instantaneous quench from a strongly interacting Mott insulator ($U\gg J$) to a weakly interacting superfluid ($J\gg U$). Our calculations are inspired by recent experiments at Munich \cite{blochlightcone}, but we consider a quench to much weaker interactions.

We calculate how density-correlations evolve following a sudden quench.  By working in the weakly interacting limit we produce analytic expressions valid for arbitrary filling factors.   In particular, for a quench to a non-interacting gas, the time dependence of the static structure factor is quite simple.  By perturbatively including the interactions, we show that density-density correlations are \textit{unaffected} to linear order in interactions. Our weak coupling calculations complement classical field studies \cite{polkovnikov, fischer, snoek} valid at large filling factors, sophisticated numerically exact approaches \cite{trotzky,flesch, rigolmtm, salvatoremtm, demlerexact, sautebd, kollath, kollathnoneq} and strong coupling theories \cite{kollathferm, altman}. Remarkably, much of the physics seen in the strong coupling calculations is already present at weak interactions. 
For example, we show that the density correlations spread ballistically \cite{cardy, kollathferm, blochlightcone, flesch, manmanalightcone}, and display damped oscillations, an effect arising purely from the underlying lattice.

\begin{figure}
\begin{picture}(100, 140)
\put(-60, -5){\includegraphics[scale=0.37]{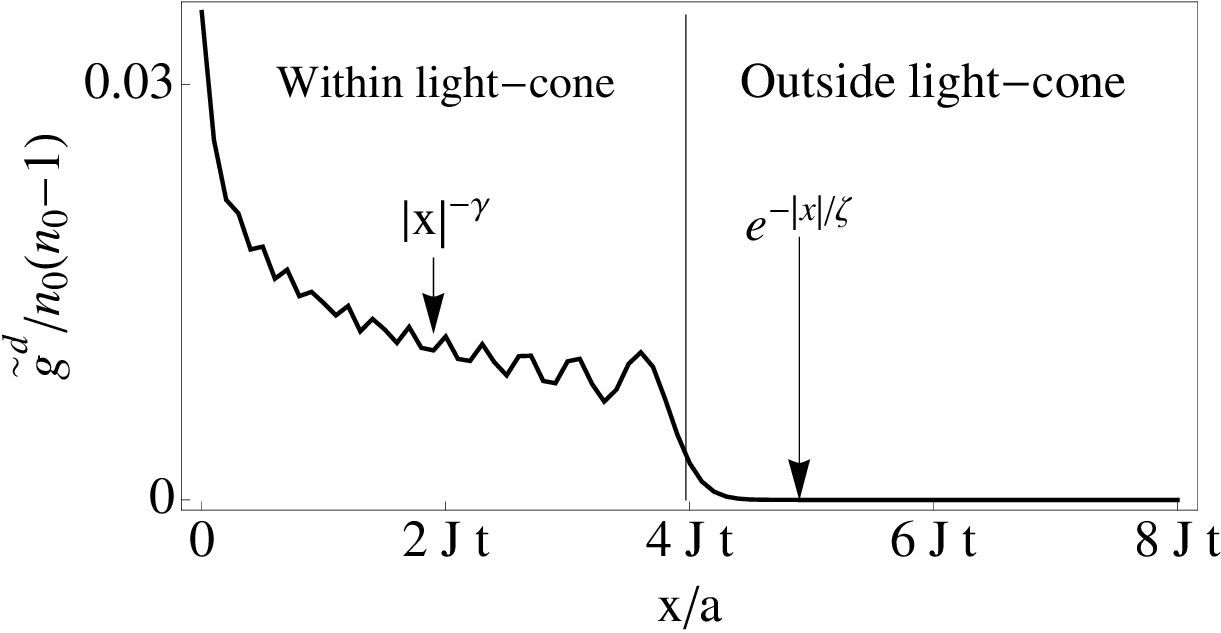}}
\end{picture}

\caption{\label{lightconevisual}\textbf{Generic structure of light-cone dynamics in $1$ dimension:} Equal-time density-density correlation function $\tilde g^{d}(\tb) = \langle a^{\dagger}_{d}(\tb)a^{\dagger}_{0}(\tb)a_{0}(\tb)a_{d}(\tb)\rangle$ (Eq.~\ref{dddyn}) plotted after some time $t$ of evolution following a quench to a \textit{non-interacting state}. The vertical line demarcates the region within the light-cone after this time. Correlations decay exponentially outside the light-cone with some correlation length $\zeta$. Within the light cone, correlations decay algebraically with some exponent $\gamma$. For the quench considered here, we numerically find $\gamma$ to be quite small, on the order of $10^{-2}$. Characteristic oscillations on the order of a lattice site are observed, arising purely from the underlying lattice band-structure.}
\end{figure}

Our studies are particularly relevant to understanding how \textit{isolated}, \textit{quantum} systems approach equilibrium \cite{weiss, kollathnoneq, trotzky, cramer, kehrein, kollathdecoh}. 
This is a relatively new area of research, primarily motivated by experiments in ultra-cold gases. 
Here we study the momentum distribution of the lattice Bose gas after a sudden quench to weak interactions. We demonstrate that in $1$D, our system does \textit{not} relax to thermal equilibrium on a timescale $t \sim J/(Un_0)^{2}$ following the quench, despite having exponentially decaying correlations in real space. However, in higher dimensions, the momentum distribution rapidly approaches a thermal distribution. For the one and two dimensional cases considered here, quasi-long range order is never established in finite time. 

\section{Formalism} 
We consider a homogeneous gas of bosons in an optical lattice described by the single-band Bose-Hubbard Hamiltonian \cite{greinerbh, fisherbh}:
\begin{equation}\label{bhham}
{\cal{H}} = -J\sum_{\langle ij \rangle}\left(a^{\dagger}_{i}a_{j} + h.c \right) +\sum_{i}\left[ \frac{U}{2}n_{i}(n_{i}-1)
- \mu n_{i}\right]
\end{equation}
where $a_{i}(\tb)$ denotes the boson annihilation operator at site $i$, $J$ denotes the hopping and $U$ the on-site repulsive interaction.  The kinetic energy sum  is over nearest neighbor pairs $\langle ij\rangle$.

The basic objects of our study are the one- and two-body density matrices; $g^{i}_{j}(\tb) = \frac{1}{i}\langle a^{\dagger}_{i}(\tb)a_{j}(\tb)\rangle$ and $g^{ij}_{kl}(\tb)= -\langle a^{\dagger}_{i}(\tb)a^{\dagger}_{j}(\tb)a_{k}(\tb)a_{l}(\tb)\rangle$. More generally we write the $n$-body density matrix as $g^{i_{1}....i_{n}}_{j_{1}...j_{n}}(\tb) = \frac{1}{i^n}\langle a^{\dagger}_{i_{i}}(\tb)....a^{\dagger}_{i_{n}}(\tb)a_{j_{n}}(t)...a_{j_{1}}(\tb)\rangle$.   In various references, these are also referred to as the $n$-body correlation functions, the 2$n$ point functions, or the equal time Green's functions.

The one- and two-body correlation functions can be readily probed in cold-atom experiments. The former is related to the momentum distribution function $g(\textbf{k}) = \frac{1}{i}\langle a^{\dagger}_{\textbf{k}}a_{\textbf{k}} \rangle = \sum_{i,j}e^{i \textbf{k} \cdotp(i-j)}g^{i}_{j}$, which is probed through bandmapping \cite{demarcoband, greinerband} or time-of-flight \cite{greinercollapse, gerbier}. The density-density correlation function $g^{ij}_{ji}$ can be measured directly using the advanced imaging techniques developed at Chicago, Harvard and Munich \cite{greinerdyn, blochquench, chengstruc}. Momentum resolved experiments such as Bragg scattering \cite{otherbragg, miyakebragg} or noise spectroscopy \cite{altmannoise, blochnoise} can be used to probe the structure factor $S(\textbf{q}) = \langle \rho^{\dagger}_{\textbf{q}}\rho_{-\textbf{q}} \rangle =  -\sum_{ij} e^{i \textbf{q}\cdotp(i - j)}g^{ij}_{ji}$, where $\rho_{\textbf{q}} = \sum_{\textbf{k}}a^{\dagger}_{\textbf{k+q}}a_{\textbf{k}}$. 

The equations of motion for the $n$-body Green's functions are constructed from the equations of motion for the operators $a_{i}(\tb)$ and $a^{\dagger}_{i}(\tb)$:
\begin{equation}\label{onebeom}
i\partial_{\tb}a_{i} = -J a_{\langle i \rangle} + Ua^{\dagger}_{i}a_{i}a_{i} - (\mu -U)a_{i}
\end{equation}
where all temporal dependence is implicit. 

For the one- and two-body Green's functions we obtain:
\begin{equation}\label{twobeom}
i\partial_{\tb}g^{i}_{j} = -J (g^{i}_{j+\langle j \rangle} - g^{i+\langle i \rangle}_{j}) -  i\hspace{0.1mm}U(g^{ii}_{ij} - g^{ij}_{jj})
\end{equation}

\begin{eqnarray}\label{fourbeom}
i\partial_{\tb}g^{ij}_{kl} = -J \left(g^{ij}_{k+\langle k \rangle}+ g^{ij}_{kl+\langle l \rangle} - g^{i+\langle i \rangle j}_{kl} - g^{ij+\langle j \rangle}_{kl}\right) \\\nonumber
- i\hspace{0.1mm}U(g^{iij}_{ikl} + g^{ijj}_{jkl}- g^{ijk}_{kkl} - g^{ijl}_{kll})
\end{eqnarray}
where the notation $\langle i \rangle$ denotes a sum over all the nearest neighbors of site $i$.   
For example, in one dimension $g^{i}_{j+\langle j \rangle}=g^{i}_{j+1}+g^{i}_{j-1}$.  In a translationally invariant system (such as the one we consider)
$g^{i}_{j+\langle j \rangle}=g^{i+\langle i\rangle}_j$, and the term proportional to $J$ in Eq.~\ref{twobeom} vanishes.

The interaction term couples the $n$-body Green's function with the $n+1$-body Green's function. The full interacting many body dynamics is described by the resulting infinite set of  coupled differential equations. 

Here we limit ourselves to the case of a shallow lattice, where interactions are weak following the quench. The single-band Bose Hubbard model is a valid description of bosons in optical lattices even for shallow lattices ($J/U \gg 1$), provided that the mean separation between the bands  is larger than the interaction energy (alternatively $V_{R}/E_{R} > 1$, where $V_{R}$ is the lattice depth). Most of the experiments are in this regime. 

Throughout this paper, we assume that the initial state at time $\tb<0$ is a homogeneous Mott insulator with $n_{0}$ bosons per site ($U = \infty$). At $\tb=0$, we suddenly quench the system to a final value of interactions $U \geq 0$ and study the subsequent evolution of the correlation functions. The dynamics is studied using a weak-coupling perturbation theory in the dimensionless parameter $U/J$ which is assumed to be small following the quench. 

Since we are interested in the weakly interacting regime, understanding the non-interacting limit is crucial \cite{kollathferm}. We first set $U=0$ and calculate the \textit{non-interacting} density-density correlation functions (Eq.~\ref{fourbeom}). We then perturbatively include the effects of $U$, determining how interactions influence the density-density correlations and the quasi-momentum redistribution in the lattice (Eq.~\ref{twobeom}). 

\section{Density-Density Correlations in $1$D}

We start by considering a one dimensional system and choose a homogeneous initial state with a density of  $n_{0}$ bosons per site. At $\tb < 0$, the sites are completely decoupled, leading to a uniform quasi-momentum distribution with magnitude $g(k) = n_{0}$. At $\tb=0$, we suddenly quench the system to a \textit{non-interacting} state $U =0$.

In the absence of interactions, there is no quasi-momentum redistribution, and the momentum occupations do not change in time. This can be easily seen by taking the Fourier transform of Eq.~\ref{twobeom}. However density-density correlations given by Eq.~\ref{fourbeom} show interesting dynamics.  

Setting $U = 0$, Eq.~\ref{fourbeom} is readily solved in Fourier space to yield $\tilde g^{pq}_{rs}(\tb) = e^{-i 2J~\tb (\cos(p)+\cos(q)-\cos(r)-\cos(s))}g^{pq}_{rs}(\tb=0)$. At $\tb = 0$, $g^{pq}_{rs}(\tb=0) = n_{0}(n_{0}-1)\delta(p+q-r-s) + n^{2}_{0}(\delta(p-s)\delta(q-r) +\delta(p-r)\delta(q-s))$. The second term generates no dynamics and produces an overall constant, which we ignore. 

In real space, the density-density correlation function then becomes:
\begin{equation}\label{dddyn}
\tilde g^{ij}_{ji}(\tb) \equiv \tilde g^{d} = n_{0}(n_{0}-1)\int_{-\pi}^{\pi}\frac{dk}{2\pi} e^{2ikd/a}J_{0}[4 J~\tb \sin(k)]^2
\end{equation}
where $\tilde g^{ij}_{ji}$ symbol is the correlation function after subtraction of the constant term, $d = i-j$, and $i^{\nu}J_{\nu}(z)  = \frac{1}{2\pi}\int^{\pi}_{-\pi}dk~e^{i(\nu k + z \cos(k))}$ is the Bessel function of first kind. A similar expression for the non-interacting limit has been also derived by Barmettler \textit{et al.} who focus on quenches to much stronger interactions ($U \geq J$) \cite{kollathferm}. 

\begin{figure}
\begin{picture}(100, 150)
\put(-80, -5){\includegraphics[scale=0.35]{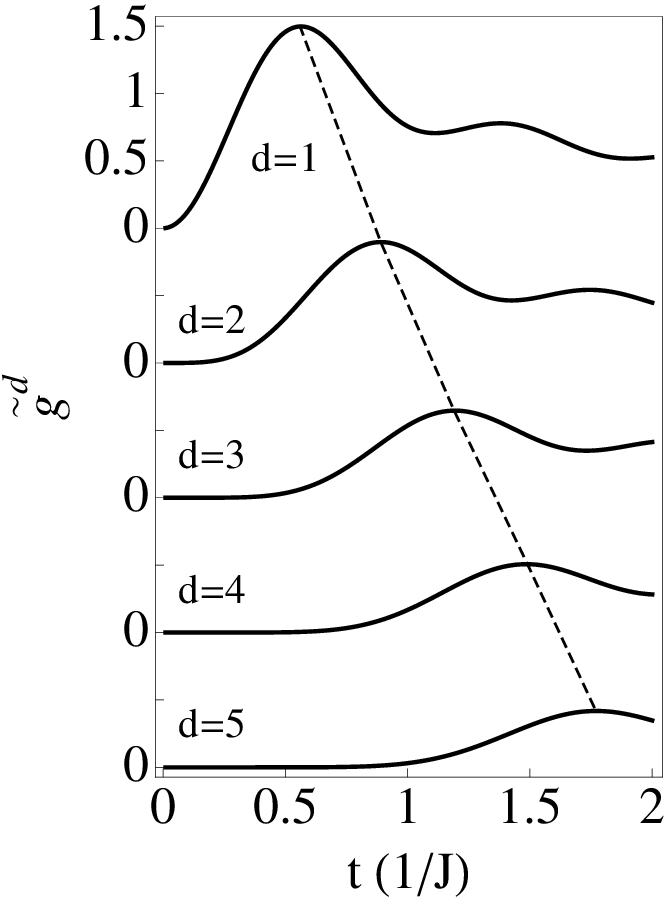}}
\put(40, -10){\includegraphics[scale=0.35]{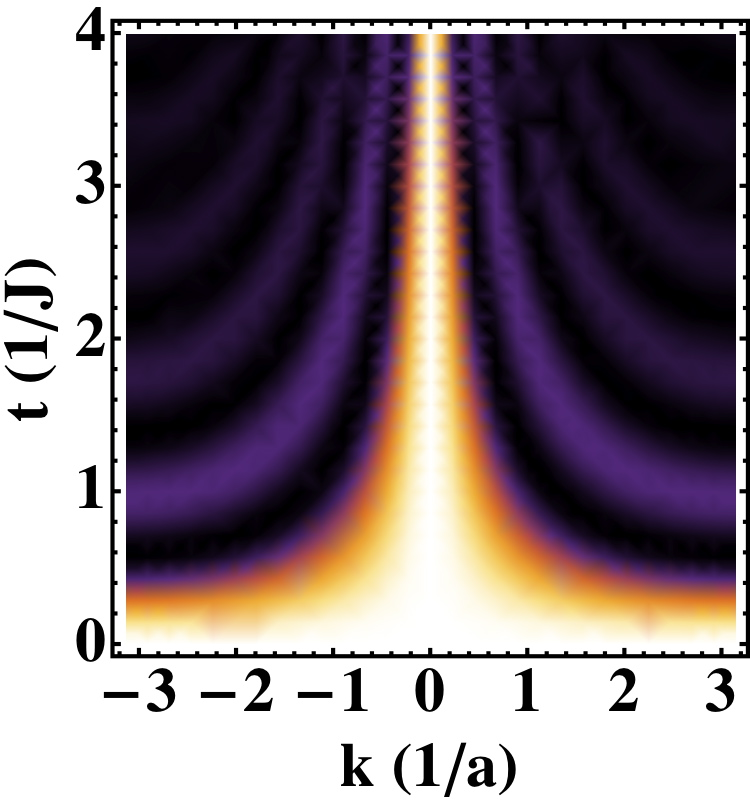}}
\end{picture}

\caption{\label{lightcone} (color online) \textbf{Light-cone evolution of density-density correlations in 1D.} (c.f. Fig. $2$, Ref.~\cite{blochlightcone}). Left:  Density-density correlation function for a homogeneous, non-interacting system $\tilde g^{ij}_{ji}(\tb) \equiv \tilde g^{d}(\tb)$ ($d = i-j$). Line shows the location of the peak in $\tilde g^{d}$ used to extract the velocity of spread of correlations. We find $v = 3.7 J a$, consistent with the spreading velocity expected for non-interacting particles.  As discussed in the main text, the structure of the density-density correlations is robust against interactions to first order in $U/J$. Right: Time-Evolution of the structure factor.  Lighter colors indicate higher intensity. At $\tb=0$, all momenta are equally occupied and $S(k)(0) = 1$ for all $k$. At intermediate times $S(k)$ shows oscillations due to interference between atomic wave-packets moving ballistically. Higher momentum contributions to $S(k)$ decay as $1/J\tb$, consistent with the linear spreading of correlations in real-space.}
\end{figure}

In Fig.~\ref{lightcone}, we plot the dynamics of the two body Green's function. As is apparent in the figure, the density correlations spread in a light-cone-like manner. One can extract a characteristic velocity associated with the ballistic spread of correlations by plotting the location of the maximum of $\tilde g^{d}$ (indicated in Fig. \ref{lightcone} by the dashed line) as a function of $d$. We obtain a velocity of $v = 3.7 J a$.  Studies by Barmettler \textit{et al.} show that this velocity has a dependence on $d$ and approaches $4Ja$ as $d \rightarrow \infty$ \cite{kollathferm}. 

We emphasize that ``light-cone dynamics" is a feature of the lattice and \textit{not} the interactions. As pointed out by Calabrese and Cardy, the initial state has very high energy ($E = 0$ in our case) compared to the ground state of the final Hamiltonian ($E_{g} = -2J$ in our case) and acts as a source for quasi-particles traveling in different directions \cite{cardy}. These matter waves carry information about correlations in the initial state. At time $\tb$ after the quench, the waves emanating from points $d = 2v\tb$ apart interfere, giving rise to an interference pattern in the density-density correlation function. In the non-interacting limit, these matter waves are simply freely propagating bosons propagating with a maximum velocity of $2Ja$ in opposite directions, giving rise to the factor of $2$ in the above expression for $d$.  Correlations decay exponentially outside the region described by the light-cone (see Fig.~\ref{lightconevisual}).

The density-density correlations for a non-interacting gas (Eq.~\ref{dddyn}) bear a striking similarity to the features observed both numerically and experimentally in the strongly interacting regime \cite{blochlightcone, kollathferm, kollath}. In fact, these features appear to be generic \cite{liebrobinson} and have also been observed in interacting Fermi systems \cite{manmanalightcone}. This similarity suggests that a similar mechanism is responsible for the build-up of correlations in the strongly interacting limit, where instead of freely propagating bosons, one has freely propagating doublon and holon pairs with a new propagation velocity. For very strong interactions, one estimates that the doublon hopping matrix element is $J_{doublon} = 2J$), and correlations propagate with a velocity $v \sim 2Ja (1 + 2) = 6 Ja$, which is consistent with the experimental and numerical findings \cite{kollathferm, blochlightcone}. 

We now briefly discuss the signatures of light-cone dynamics in momentum space. In Fig.~\ref{lightcone}, we also plot the structure factor obtained by taking the Fourier transform of the density-density correlation function. This can simply be read off from Eq.~\ref{dddyn} as $S(q)(\tb) = n_{0}(n_{0}-1) J_{0}[4 J~\tb \sin(q/2)]^{2}$. At $\tb=0$, the structure factor is a constant as all momentum states are equally occupied. As the system begins to develop correlations between neighboring sites, the structure factor shows periodic oscillations whose amplitude decays in time. Using the asymptotic behavior of the Bessel function $J_0(z)\sim (2/\pi z)^{1/2}\cos(z-\pi/4))$ as $z\to\infty$, we find that for long times the oscillations have period $\tau_{osc} =  \pi/[4J\sin(k/2)]$.  At long times, the correlations are found to decay to steady state values as $1/J\tb$. In $k$ space, the envelope of the structure factor decays as $1/k$. These features can be readily accessed in experiments. 

The long time behavior of correlations however is very different in the non-interacting and strongly interacting limit. In contrast with the rather slow decay of correlations for the quench to $U=0$, density-density correlations appear to decay rapidly in the strongly interacting case \cite{natubogquench, blochlightcone, kollathferm}. The mechanism for the decay contains information about the nature of the quasi-particles and their interactions and merits further study.  

We now consider a quench to a weakly interacting final state $U/J \ll 1$ and compute the effect on the density-density correlation functions to first order in perturbation theory. Interestingly, we find that the non-interacting density-density correlations are completely unaffected. 

In order to calculate $g^{pq}_{rs}$ to first order in interactions, we Fourier transform Eq.~\ref{fourbeom} and assume that the three-body correlator $g^{pqr}_{suv}$ evolves freely as it would for a non-interacting system. We then substitute the expression for the three-body correlation function into Eq.~\ref{twobeom} and obtain the two-body correlation function in Fourier space (details are supplied in Appendix A):
\begin{widetext}
\begin{eqnarray}\label{g2dynfinal}
g^{pq}_{rs}(\tb) = \Biggl(g^{pq}_{rs}(0) - i\frac{Un^{2}_{0}(n_{0}-1)}{J} \times \int^{\tb}_{0}d\tau \delta_{p+q-r-s}\sum_{\delta}J_{\delta}(\tau)^{2} \Bigl(i^{2\delta}e^{i\tau(\cos{p}+\cos{q})}e^{-i\delta(q+p)} -\hspace{30mm}\\\nonumber  i^{-2\delta}e^{-i\tau(\cos{r}+\cos{s})}e^{i\delta(r+s)}\Bigr) - i\frac{Un_{0}(n_{0}-1)(n_{0}-2)}{J} \times \int^{\tb}_{0}d\tau \delta_{p+q-r-s}\sum_{\delta}J_{\delta}(-\tau)^{3} \Bigl(i^{\delta}(e^{i\tau\cos{p}}e^{-i\delta p} +e^{i\tau\cos{q}}e^{-i\delta q})-\\\nonumber  i^{-\delta}(e^{-i\tau\cos{r}}e^{i\delta r}+e^{-i\tau\cos{s}}e^{i\delta s})\Bigr)\Biggr)\times  e^{-i\tb(\cos{p} + \cos{q} -\cos{r}-\cos{s})}
\end{eqnarray}
\end{widetext}
The first term in the brackets is the non-interacting two-point correlation function which now acquires a time and momentum dependent correction of order $U/J$ from the three-body terms (Eq.\ref{fourbeom}).

Taking the Fourier transform of the above expression, one finds that $g^{ij}_{ji}$ is completely unaffected to linear order in $U/J$, for any filling.  Our calculations imply that for a quench to the weakly interacting regime, $g^{ij}_{ji}$ scales as $g^{ij}_{ji}(\tb) \sim g^{(0) ij}_{ji}(\tb) + {\cal{O}}(Un/J)^{2}$, where $g^{(0) ij}_{ji}(\tb)$ is the non-interacting density-density correlation function calculated above.



Different behavior is found when the initial state is a weakly interacting superfluid \cite{natubogquench}. In this case, following the quench, the density-density correlation function to leading order is proportional to $U~n_{0}n_{ex}$ where $n_{0}$ is the condensate density and $n_{ex}$ is the density of quasi-particle excitations out of the condensate.

\begin{figure}
\begin{picture}(100, 270)
\put(-75, 155){\includegraphics[scale=0.55]{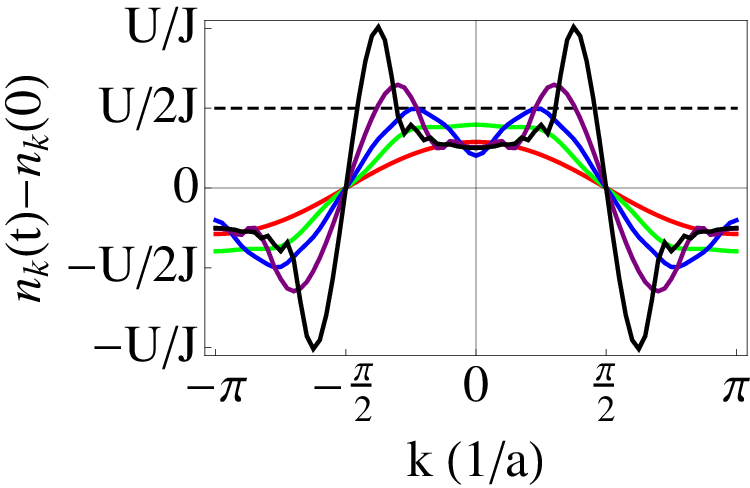}}
\put(-70, -5){\includegraphics[scale=0.35]{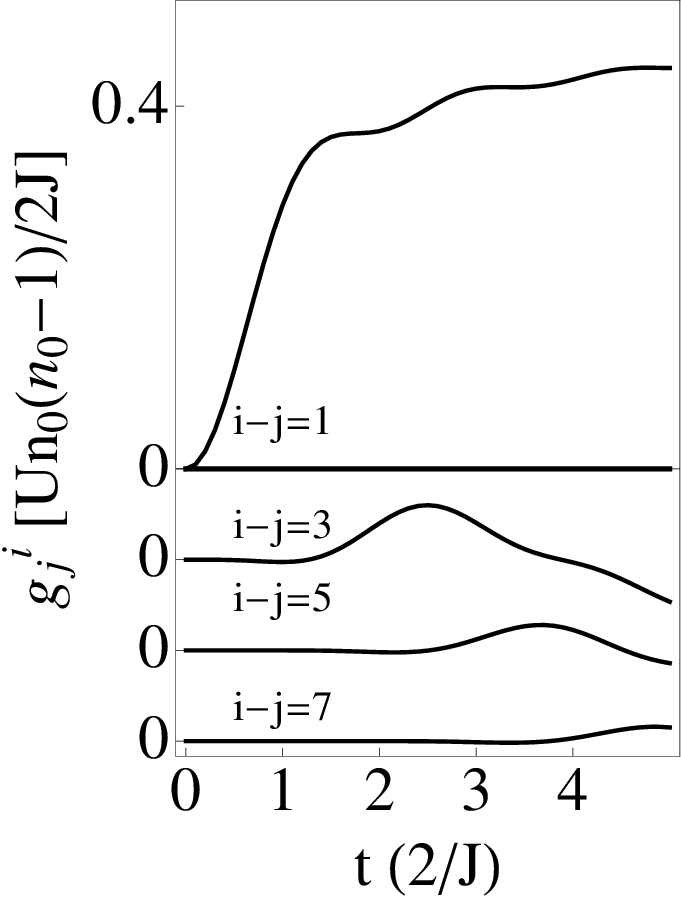}}
\put(50, 35){\includegraphics[scale=0.34]{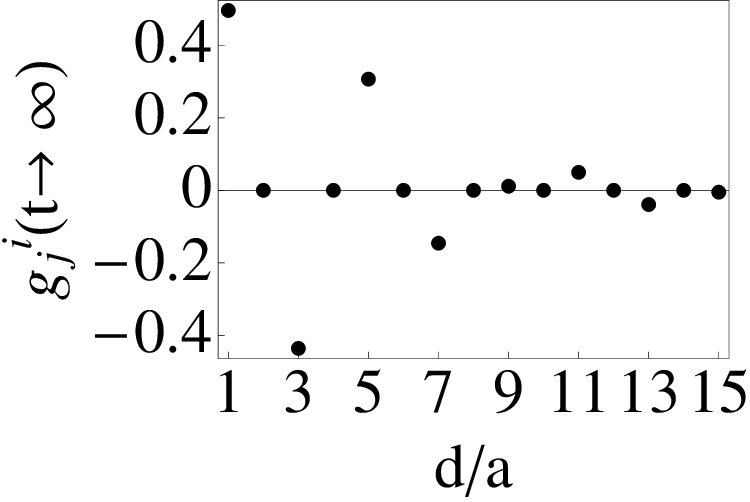}}
\end{picture}

\caption{\label{tofev} (Color online) \textbf{Coherent redistribution of quasi-momentum to linear order in $U/J$ in 1D.} Top: Redistribution of quasi-momentum ($n_{k} = \langle a^{\dagger}_{k}a_{k} \rangle$) at times $\tb=0$ (thick, dashed), $\tb =0.5/J$ (red), $2.5/J$ (green), $5/J$ (blue), $12.5/J$ (purple), $25/J$ (black) obtained by integrating Eq.~(\ref{qmtmev}). At short times, we find a coherent transfer of quasi-momentum from high momentum states to low momentum states. At longer times,  we find a pile-up of particles near $k = \pm\pi/2$.  Bottom (Left): Spatial evolution of the one-body density matrix $g^{i}_{j}(\tb)$ for different values of $d=i-j$. Short range correlations rapidly saturate while longer range correlations take time to develop. (Right): Correlations $g^{i}_{j} - n_{0}$ measured in units of $Un_{0}(n_{0}-1)/2J$ in real space as a function of $d = i-j$ at long times $t = 100/J$. Correlations vanish if $d$ is even. Appreciable long range correlations never develop even on long times. }
\end{figure}

\section{Momentum distribution in $1$D}
As in the case of the density-density correlations, we calculate the momentum distribution perturbatively in the final (dimensionless) interaction strength $U/J$. In the absence of interactions ($U=0$), there is no momentum redistribution. Thus we must take $U\neq0$ after the quench in Eq.~\ref{twobeom}. 

To leading order in the interaction strength $U$, we solve Eq.~\ref{twobeom} by replacing the two body correlator
$g_{ij}^{kl}(\tb)$ with the noninteracting result in Eq.~\ref{dddyn}.  
We find that the occupation numbers obey:

\begin{eqnarray}\label{qmtmev}
\partial_{\tb}g_{q}(\tb) = \frac{U n_{0}(n_{0}-1)}{2J}\sum^{\infty}_{k= -\infty}J_{k}(-\tb)J^{2}_{k}(\tb) \times  \hspace{5.5mm} \\\nonumber
\left(i^{-k}e^{i(qk - \tb \cos(q))} - i^{k}e^{-i(qk - \tb \cos(q))}\right) 
\end{eqnarray}
where we have normalized time in units of $1/2J$. Note that to first order in $U/J$, interactions merely shift the magnitude of the momentum distribution. The right-hand-side of Eq.~\ref{qmtmev} is invariant under the inversion $q \rightarrow -q$ but switches sign under the transformation $q \rightarrow \pi-q$. This implies that $q=\pm\pi/2$ is a stationary point and states at $q=\pi/2$ have no dynamics. In real space, this symmetry implies $g^i_j=0$ if $d=|i-j|$ is even.




To lowest order in interactions, we assume that the two-body correlation function behaves as if interactions are absent, \textit{i.e} every momentum state evolves independently $g^{pq}_{rs} \sim \delta(p+q-r-s)e^{-i2J\tb(\cos(p)+\cos(q)-\cos(r)-\cos(s))}$, while conserving total momentum. States at $q = \pm\pi/2$ do not evolve, as $\cos(q)$ vanishes here.  


In Fig.~\ref{tofev} we plot the evolution of the quasi-momentum states obtained by integrating  Eq.~\ref{qmtmev}. At $\tb=0$, all momentum states are equally occupied. At short times following the quench, quasi-momentum states explore the band and the low momentum occupation begins to grow. 
At intermediate times, the momentum distribution develops peak-like features which migrate towards the stationary points $q = \pm\pi/2$. Expanding Eq.~\ref{qmtmev} near $q = \pi/2$, one finds that the slope of the momentum distribution near $\pi/2$ grows as $(\tb/J)^{2/3}$. 

At long times, the rate of momentum redistribution slows down, and the system settles into a more or less steady state with a relatively flat quasi-momentum profile near $q=0$, and sharp peaks near $q = \pm\pi/2$. Owing to this spectral feature, in real space only $g^{i}_j$ (with $d=|i-j|$ odd) are appreciable at long times.

The Fourier transform of the momentum distribution reveals the dynamics of the one-body density matrix, which is plotted in Fig.~\ref{tofev}. At short times the single-particle correlations spread in a manner similar to the density correlations.  Local correlations are rapidly established on a time of order $J^{-1}$.  Long range order, however, requires communication between widely separated sites and take longer to develop. 

As correlations can develop at best linearly in time, infinite range order is not found at any finite time. This is evidenced in Fig.~\ref{tofev} (bottom-right) where the one-body density matrix is plotted as a function of the separation $d = i-j$ between sites at long times. The envelope of the one-body density matrix (for odd sites) is found to decay \textit{exponentially} indicating an absence of any long range order. 

Although the system reached a steady state, with exponentially decaying correlations in real space, the momentum distribution in Fig.~\ref{tofev} is distinctly ``athermal". We attribute this to the fact that to first order in $U/J$ the evolution conserves the occupation of quasi-momentum at $q = \pm\pi/2$.

It is then natural to ask whether this momentum distribution will survive when particles are allowed to scatter to and from $q=\pm\pi/2$. These effects first enter at order $(U/J)^{2}$, and are considered below. By substituting the first order result for the two-point function $g^{pq}_{rs}$ (Eq.~\ref{g2dynfinal}) into the expression for the momentum distribution Eq.~\ref{twobeom}, we can evaluate the dynamics of the momentum distribution to second order in $U/J$. 


The full expression for the momentum distribution upon inclusion of the second order terms reads:
\begin{widetext}
\begin{eqnarray}\label{momdistfinal}
\partial_{\tb}n_{q}  = -i\frac{Un_{0}(n_{0}-1)}{2J}\Bigl[\sum_{k}J^{2}_{k}(\tb)J_{k}(-\tb)\Bigl(i^{-k}e^{i(kq - \tb\cos{q})}- i^{k}e^{-i(kq - \tb\cos{q})}\Bigr)\Bigr]+\hspace{30mm} \\\nonumber \frac{(Un_{0})^{2}(n_{0}-1)}{J^{2}}{\cal{R}}\Biggl[\sum_{k, \delta}\int^{\tb}_{0}d\tau J^{2}_{\delta}(\tau)J_{-k}(\tb) \Bigl(J_{k-\delta}(\tau-\tb) +\frac{n_{0}-2}{2n_{0}} J_{-\delta}(\tau)J_{k}(-\tb)\Bigr)\Bigl[i^{k}J_{k-\delta}(\tau-\tb)e^{-i(k q - \tb \cos{q})}  -\\\nonumber i^{k-\delta}J_{k}(-\tb)e^{-i((k-\delta)q + (\tau-\tb)\cos{q})}\Bigr]\Biggr]
\end{eqnarray}
\end{widetext}
where ${\cal{R}}$ denotes the real part of the expression. Some of the details of the calculation are presented in Appendix B.

The first term in the right hand side of Eq.~\ref{momdistfinal} is simply the first order result, rewritten. 

The second term, proportional to $(U/J)^{2}$ has two contributions:  The term proportional to $n^{2}_{0}(n_{0}-1)$ represents the scattering of two-particles and is the dominant process at this order. In addition, there is a sub-leading contribution (which has an additional factor of $J_{k}$ in Eq.~\ref{momdistfinal}) which arises due to scattering of \textit{three} particles.

In Appendix B, we discuss both these terms and their effect on the momentum distribution independently. We find that unlike the first order result which was anti-symmetric about $q = \pm\pi/2$, both the ${\cal{O}}(U/J)^{2}$ terms give rise to a distribution that is symmetric about $q=\pm\pi/2$. The term proportional to $n_{0}(n_{0}-1)(n_{0}-2)\delta(p+q+r-s-u-v)$ in Eq.~\ref{threegf} tends to \textit{decrease} the occupation of momentum states near $q=\pi/2$, while terms like $n^{2}_{0}(n_{0}-1)\delta_{ps}\delta_{q+r-u-v}$ in Eq.~\ref{threegf} tends to \textit{increase} the occupation near $q=\pi/2$. To quadratic order in perturbation theory, this term dominates over the former, ultimately \textit{enhancing} the peak-like features seen at finite momentum. Evolving the system for longer times the momentum occupation develops \textit{symmetric} peaks about $q = \pm\pi/2$. 

\begin{figure}
\begin{picture}(100, 200)
\put(-40, 85){\includegraphics[scale=0.35]{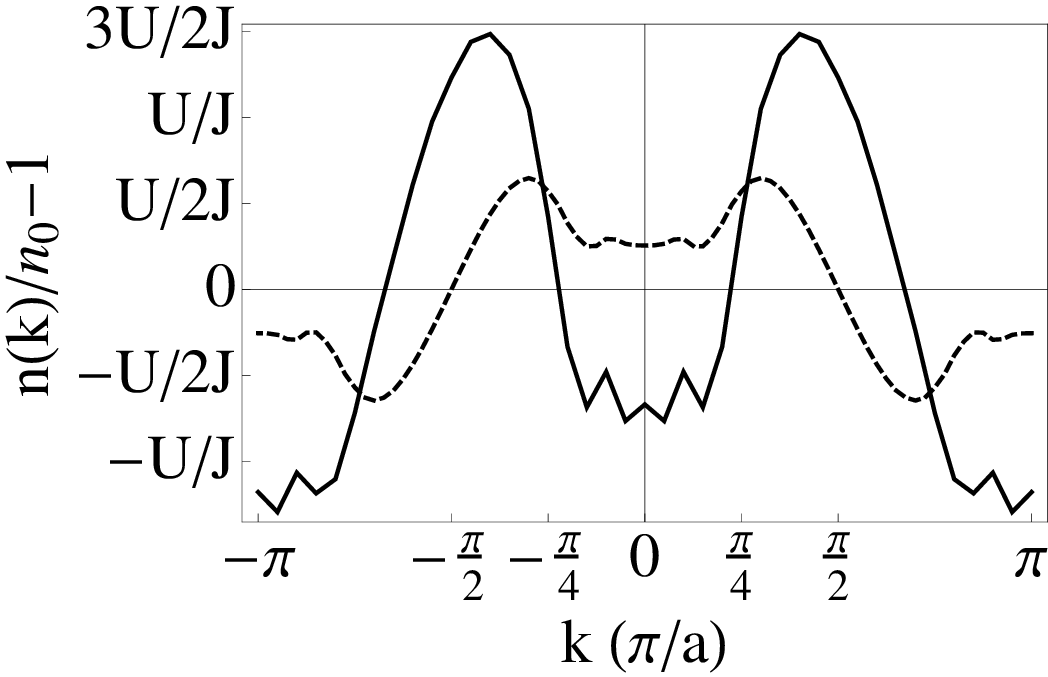}}
\put(-20, -8){\includegraphics[scale=0.375]{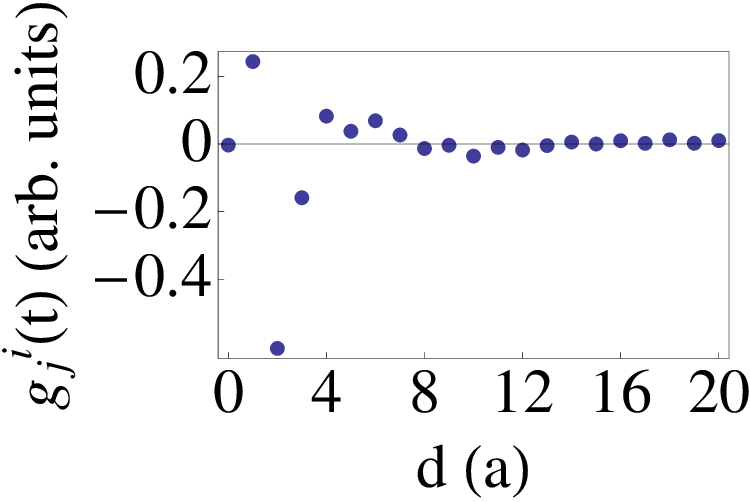}}
\end{picture}

\caption{\label{collfig} (Color online) \textbf{Redistribution of quasi-momentum to ${\cal{O}}(U/J)^2$ in 1D.} Top: Redistribution of quasi-momentum ($n_{k} = \langle a^{\dagger}_{k}a_{k} \rangle$) at time $\tb=12.5/J$ (thick, solid), compared with the first order result at the same time (dashed). The interaction strength has been chosen to be $Un_{0}/J = 0.3$ to highlight the features of the second order calculation. On times $\tb \sim 10/J$, the occupation of quasi-momentum near $q=\pm\pi/2$ grows in time, suppressing the occupation at zero momentum. Our calculations are valid for times $\tb \sim J/U^2$. (Bottom): Evolution of the one-body density matrix after time $\tb =10/J$. To quadratic order in the interactions, correlations build up between even sites. The envelope of the correlation function decays exponentially, indicating the absence of quasi-long range order.}
\end{figure}

In Fig.~\ref{collfig}, we plot the momentum distribution upon inclusion of the quadratic terms. We attribute the appearance of peaks at  $q=\pi/2$ to the restricted phase space available for scattering in $1$D. Near $q=\pi/2$, the dispersion becomes linear and the constraints of momentum and energy conservation relax into a single constraint. One may expect therefore that the bulk of the two particle scattering occurs near these points. Unlike fermions, Bose statistics tends to \textit{enhances} the probability of scattering into states that are already occupied, thus leading to an enhancement of the peaks over time. 

We emphasize however that our approach only captures the initial stages of equilibration. A full treatment of thermalization should take multiple scattering processes into account and is beyond the scope of this paper. In the Appendix we show that scattering of three particles tends to suppress the occupation near $q=\pi/2$. These processes will become important on times $\tb \sim J^{2}/U^{3}$, and may eventually drive the system to a thermal distribution. 


The structure near $q=\pm\pi/2$ in Fig.~\ref{collfig} is reminiscent of the peaks seen in simulations of expanding $1$D interacting bosons by Rigol and Muramatsu \cite{rigolmtm} and subsequently by Rodriguez \textit{et al.} \cite{salvatoremtm}. Our calculation which is valid for times $t < J/U^{2}$, finds a similar suppression in the momentum occupation at $k=0$.  Taking the Fourier transform of the momentum distribution, we find that the one-body density matrix now develops correlations between sites separated by even lattice spacings. However at long distances, correlations still decay exponentially, and long range order is not observed. Thus the peaks seen in the momentum distribution in our case \textit{do not} correspond to a quasi-condensate. 

Our calculations are similar in spirit to the interaction quench considered by Mo\"eckel and Kehrein in the fermionic Hubbard model \cite{kehrein}. The picture they develop is that the system shows an initial build-up of correlations, reaching a non-thermal steady state on intermediate times, and an eventual approach to equilibrium on much longer timescales. Our calculations point to a similar picture for quenches in lattice bosons. It will be extremely interesting to understand why this picture is generic.  


\section{Two dimensions} We now generalize our results to higher dimensions. Concretely, we consider the case of a two-dimensional square lattice, initially containing $n_{0}$ particles per site, and investigate the dynamics following a sudden reduction of  the lattice depth to the weakly interacting limit. 

\begin{figure}
\begin{picture}(100, 110)
\put(-35, -10){\includegraphics[scale=0.45]{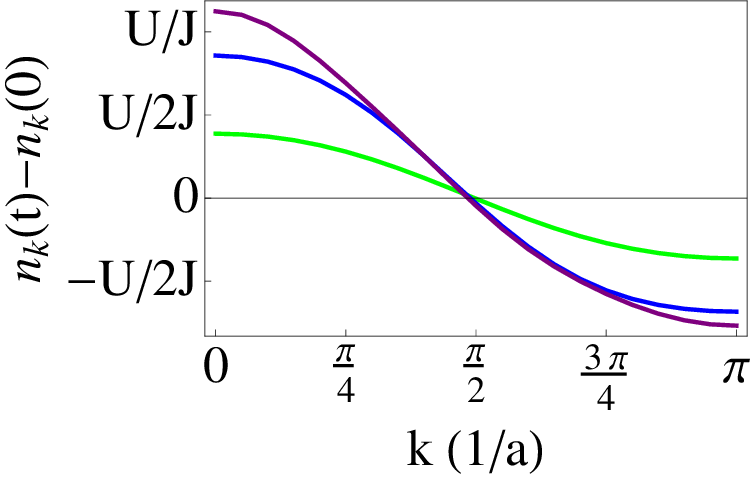}}
\end{picture}
\caption{\label{2ddensdist}\textbf{Rapid equilibration of momentum distribution in two-dimensions} Momentum distribution along the $\{\pi, \pi\}$ vector obtained by integrating Eq.~(\ref{twobeom}) in two-dimensions assuming an initially uniform distribution. The blue, green and purple curves correspond to times $\tb = 0.25/J$, $\tb = 0.5/J$ and $\tb = 1/J$ respectively. In contrast to the one-dimensional case, the distribution evolves rapidly to a broad peak at $k=0$, with no further dynamics.}
\end{figure}

Since the ``light-cone" effect arises primarily due to the bounded lattice spectrum and not the interactions, one expects it to persist in higher dimensions as well. 

We now calculate the density-density correlations for a quench to $U=0$ in $2$D. Repeating our $1$D arguments in one higher dimensions, one immediately finds the structure factor evolves according to 
$S(q_{x}, q_{y})(\tb) = n_{0}(n_{0}-1)J_{0}[4 J~\tb \sin(q_{x}/2)]^{2}J_{0}[4 J~\tb \sin(q_{y}/2)]^{2}$. As in the one-dimensional case, by taking the Fourier transform of the above expression one finds that the density-density correlations evolve in a manner identical to the one-dimensional case, with a characteristic velocity that now depends on direction.  At long times, correlations decay with a power law $1/\tb^{2}$ (as opposed to the $1/\tb$ decay in one-dimension). After a time $\tb$, correlations spread over a volume $\sim v^2\tb^2$ where $v$ is twice the characteristic velocity of an free particle (for example, $v\sim 4 \sqrt{2}Ja$ along the $\{\pi, \pi\}$ wave-vector).  

In analogy with the $1$D calculations, we calculate the momentum distribution following a quench in $2$D, finding dramatic differences. Including interactions perturbatively to order ${\cal{O}}(U/J)^2$, we plot in Fig.~\ref{2ddensdist}, the momentum distribution along $\{\pi, \pi\}$.   
At long times the distribution is characterized by a broad peak centered around $k=0$. The presence of a broad peak indicates that only short range correlations are developed, and the absence of any long range order (either true long range order or algebraic). 

An important difference between the one and two-dimensional results is the \textit{timescale} for momentum distribution. While the $1$D distribution continues to evolve on times $t \sim 50/J$, the $2$D momentum distribution reaches a steady state much faster. 
This is due to the rapid decay of density-density correlations in higher dimensions, which drive the redistribution of quasi-momentum. Our findings are consistent with numerical calculations by Sau, Wang and Sarma \cite{sautebd} who consider quenches to much stronger interactions $U \sim 2J$, finding that the final momentum distribution rapidly becomes thermal.  

\section{Summary} By considering the dynamics of lattice bosons following a quench to a weakly interacting final state, we have explored how correlations develop in a many-body system.  Our analytic work complements the large body of numerical work on this subject by working in a regime where numerics is prohibitive due to the large Hilbert space needed to accurately capture the dynamics. 

Surprisingly, much of the behavior seen in the strongly interacting system is already present for weak interactions.  For example, we find that correlations develop in a manner similar to those seen in experiments \cite{blochlightcone}. We emphasize that these features are merely \textit{lattice} effects and should not be attributed to strong interactions. Numerical studies have also found that the light-cone behavior is generic to a wide range of interaction strengths \cite{manmanalightcone, kollathferm}. We have also shown that the these features in the density density correlation function are robust to first order in perturbation theory in the interactions. 


In addition, we have studied how quasi-momentum states evolve following the quench. Surprisingly we find that for a quench to weak interactions, the quasi-momentum distribution develops peaks at finite momentum that migrate to $k = \pm\pi/2$ over time. In real space this implies correlations between sites separated by odd lattice spacings. It will be extremely interesting to observe this signature experimentally or in numerical simulations. By working to second order in $U/J$ we show that these peaks are robust on times $t \lesssim J/U^{2}$. 

The nature of the one-body density matrix is directly related to understanding whether the system develops long-range order after a quench. Over a decade ago there was a large body of work asking analogous questions with thermal quenches \cite{ao}. The picture they developed was one of nucleation and subsequent coarsening. Similar physics is expected in the quantum case \cite{mukundquench, spielmancoarsening}.  Here we show that for a quench from the insulating phase, long range order is not established after a finite time (either algebraic or true), and the one-body density matrix decays exponentially in real space. Nonetheless we find a highly non-trivial momentum distribution in $1$D, indicating that the dynamics is non-ergodic. 


\section{Future Directions for Theory and Experiment}

We conclude this paper with a discussion of what in our view constitute important future directions for theory and experiment. A key question to understand is how properties of the initial and final state after the quench are reflected in the short and long time dynamics of correlations \cite{natubogquench, chengsakh, blochlightcone, trotzky}. For example, a key difference between our calculations and the experimental and numerical findings is the rapid decay of correlations in the latter case. It will be extremely interesting to study whether one can extract properties of the excitation spectrum and quasi-particle decay rates from this long time behavior. An important limitation of state of the art numerical methods is that they are restricted to one-dimension or small system sizes in higher dimensions. Mean-field or Boltzmann equation type approaches that take into account correlations in the initial state may be able to shed light on the dynamics of quasi-momentum in higher dimensions.

Here we have shown that non-trivial dynamics occurs even for quenches to weak interactions \cite{natubogquench}. It will be extremely interesting to explore this parameter regime experimentally. In particular the momentum distribution after a quench can be readily obtained by time-of-flight or bandmapping.  A major advantage of experiments is that they can be performed in higher dimensions, where theory is largely restricted to mean-field type approaches that typically do not capture correlations fully \cite{natudemarco}. 

An important question for both theoretical and experimental consideration is to understand whether non-integrable systems generically approach equilibrium in a three-step manner \cite{kehrein}: on short times, the system is effectively ``collisionless" and supports freely propagating quasi-particles bearing information about the initial state; on intermediate timescales it approaches a non-thermal but steady state due to interference and dephasing between these quasi-particles and on long times, the system loses memory of its initial state, and ultimately approaches equilibrium driven largely by collisions between low energy degrees of freedom.  

We hope that future experiments along these lines will be able to determine the nature of the final state after such a quench and settle questions regarding the emergence of long range order and thermalization in isolated quantum systems.




\section{Acknowledgements} This work is supported by a grant from the Army Research Office with funding from the DARPA OLE program and was partially completed at the Aspen Center for Theoretical Physics and the Kavli Institute for Theoretical Physics (KITP), supported by NSF grant Nos. PHY-$1066293$ and PHY $1125915$. SN would also like to thank the organizers and participants of the KITP program entitled \textit{Quantum Dynamics in Far from Equilibrium Thermally Isolated systems} for numerous engaging and stimulating discussions. In particular, SN would like to thank David Huse and Marco Schiro for their insights. 

\appendix

\section{Dynamics of the two-body correlation function to ${\cal{O}}(U/J)$}
\numberwithin{equation}{section}
\renewcommand{\theequation}{A-\arabic{equation}}

Here we discuss the derivation of Eq.~\ref{g2dynfinal} in the main text. In Appendix B, we will use this formula to derive the equations of motion governing the dynamics of the momentum distribution to order ${\cal{O}}(U/J)^2$. 

We start by Fourier transforming Eq.~\ref{fourbeom} to obtain:
\begin{widetext}
\begin{eqnarray}\label{fourbeomf}
\Bigl(i\partial_{\tb} - 2J(\cos{p}+\cos{q}-\cos{r}-\cos{s})\Bigr)g^{pq}_{rs}  = -iU \int dxdydz \Bigl(g^{xyq}_{zrs}\delta(x+y-z-p) +g^{pxy}_{zrs}\delta(x+y-z-q) \\\nonumber -g^{pqx}_{yzs}\delta(y-z-x-r) - g^{pqx}_{yzr}\delta(y-z-x-s)\Bigr)
\end{eqnarray}
\end{widetext}
where the n-body Green's function in real space is given by:
\begin{equation}
g^{i_{1}....i_{n}}_{j_{1}...j_{n}}(\tb) = \frac{1}{i^n}\langle a^{\dagger}_{i_{i}}(\tb)....a^{\dagger}_{i_{n}}(\tb)a_{j_{n}}(\tb)...a_{j_{1}}(\tb)\rangle
\end{equation}

Assuming a homogeneous initial state of $n_{0}$ bosons per site, we expand the \textit{three} body correlation function as: 
\begin{eqnarray}\label{threegf}
-i g^{pqr}_{suv}(\tb=0) = n_{0}(n_{0}-1)(n_{0}-2) \delta_{p+q+r-s-u-v} \hspace{7mm} \\\nonumber + n^2_{0}(n_{0}-1)(\delta_{ps}\delta_{q+r-u-v} + ...) + n_{0}^{3}(\delta_{ps}\delta_{qu}\delta_{rv} + ...) 
\end{eqnarray}
where the $...$ indicate all possible terms of that type. For example, for the term proportional to $n^{2}_{0}(n_{0}-1)$, there are $9$ such possibilities. 

We now assume that the three body correlator evolves as if the system were non-interacting:
\begin{eqnarray}\label{threegfdyn}
g^{pqr}_{suv}(\tb) = g^{pqr}_{suv}(\tb=0)\times \hspace{30mm}\\\nonumber e^{-2iJ\tb (\cos(p) +\cos(q)+\cos(r)-\cos(s)-\cos(u)-\cos(v))} 
\end{eqnarray}


Although the expression for the three-body correlation function has many terms, the calculation rapidly simplifies. First note that the terms of the form $\delta_{ps}\delta_{qu}\delta_{rv}$ produce no dynamics and give rise to constants that cancel out when substituted into Eq.~\ref{fourbeomf}. 

The terms proportional to $n_{0}(n_{0}-1)$ have to be considered carefully. The $9$ terms that add up to $g^{xyq}_{zrs}$ are:
Expanding the terms one finds:
\begin{widetext}
\begin{equation}\label{g3bracket}
g^{xyq}_{zrs}(\tb=0) = (\delta_{xz}\delta_{yqrs} + \delta_{yz}\delta_{xqrs}) + (\delta_{xr}\delta_{yqzs} + \delta_{xs}\delta_{yqzr} +   \delta_{yr}\delta_{xqzs} + \delta_{ys}\delta_{yqzr}) + \delta_{qz}\delta_{xyrs} + (\delta_{qr}\delta_{xyzs} + \delta_{qs}\delta_{xyzr})
\end{equation}
\end{widetext}
where the brackets indicate terms which yield similar forms upon integration. Also the delta-function $\delta_{abcd}$ is short-hand for $\delta_{a+b-c-d}$. 

By considering each of the brackets separately for the $4$ terms in the right hand side of Eq.~\ref{fourbeomf}, one finds that only the term $\delta_{qz}\delta_{xyrs}$ yields a non-zero result. Therefore from terms proportional to $n^{2}_{0}(n_{0}-1)$ we obtain the equation:
\begin{widetext}
\begin{eqnarray}\label{g2dyn1}
\Bigl(i\partial_{\tb} - 2J(\cos{p}+\cos{q}-\cos{r}-\cos{s})\Bigr)g^{pq}_{rs}  =- iUn^{2}_{0}(n_{0}-1) \delta_{p+q-r-s}\sum_{\delta}J_{\delta}(\tb)^{2} \times\\\nonumber \Bigl(i^{2\delta}e^{i\tb(\cos{r}+\cos{s})}e^{-i\delta(q+p)} - i^{-2\delta}e^{-i\tb(\cos{p}+\cos{q})}e^{i\delta(r+s)}\Bigr)
\end{eqnarray} 
\end{widetext}

A similar calculation for the term proportional to $n_{0}(n_{0}-1)(n_{0}-2)$ yields:
\begin{widetext}
\begin{eqnarray}\label{g2dyn2}
\Bigl(i\partial_{\tb} - 2J(\cos{p}+\cos{q}-\cos{r}-\cos{s})\Bigr)g^{pq}_{rs}  = - iUn_{0}(n_{0}-1)(n_{0}-2) \delta_{p+q-r-s}\sum_{\delta}J_{\delta}(-\tb)^{3} \times  \hspace{10mm} \\\nonumber \Bigl(i^{\delta}(e^{-i\delta p}e^{-i\tb(\cos{q}-\cos{r}-\cos{s})}  e^{-i\delta q}e^{-i\tb(\cos{p}-\cos{r}-\cos{s})}) - i^{-\delta}(e^{i\delta r}e^{-i\tb(\cos{p}+\cos{q}-\cos{s})}+e^{-i\delta s}e^{-i\tb(\cos{p}+\cos{q}-\cos{r})})
\end{eqnarray} 
\end{widetext}

These equations can be solved by first making a transformation to rotating coordinates to eliminate the $2J$ term on the left, integrating the resulting equation and then transforming back. Adding Eq.~\ref{g2dyn1} and \ref{g2dyn2} and performing this operation yields Eq.~\ref{g2dynfinal}. 

\section{Dynamics of Momentum Distribution to ${\cal{O}}(U/J)^2$}
\numberwithin{equation}{section}
\renewcommand{\theequation}{B-\arabic{equation}}

\begin{figure}
\begin{picture}(100, 200)
\put(-45, 95){\includegraphics[scale=0.4]{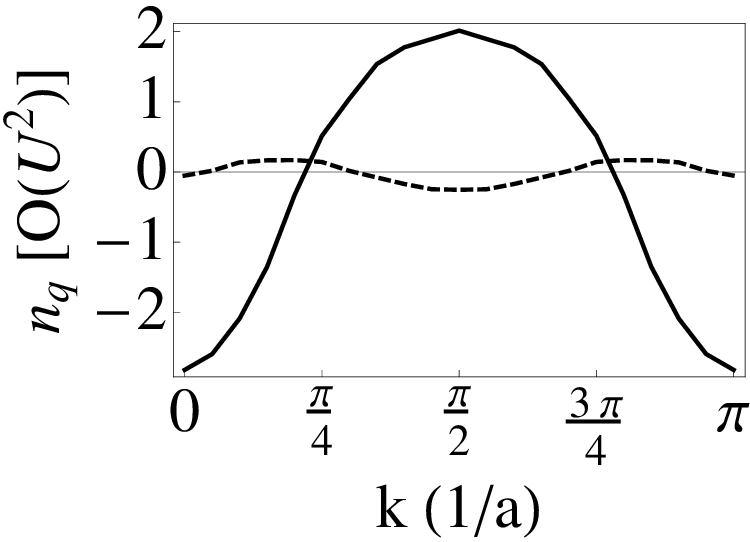}}
\put(-40, -12){\includegraphics[scale=0.4]{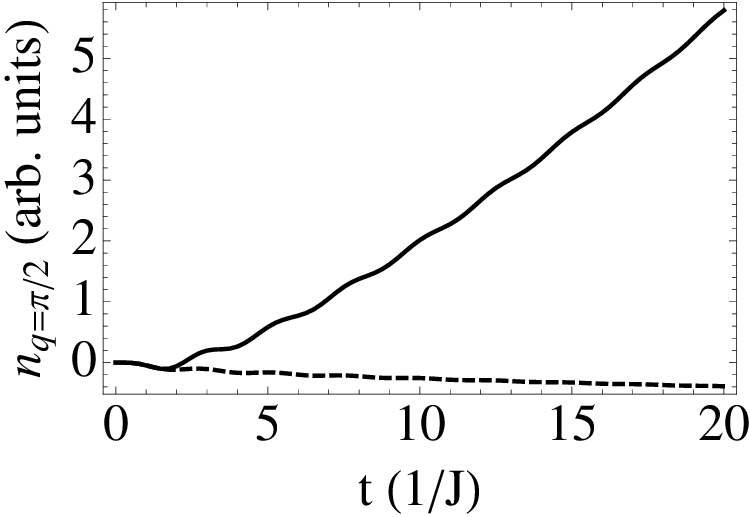}}
\end{picture}
\caption{\label{momdistquad}\textbf{Role of Quadratic corrections in the evolution of $n_{q}$} Top: Solid line is the evolution of the momentum distribution (Eq.\ref{momdistfinal}) at time $\tb = 10/J$, arising purely from  terms in Eqs.~\ref{mom1} and \ref{mom2}, showing an enhancement in the peaks at $\pi/2$. Dashed line is the evolution of the momentum distribution at time $t= 10/J$ arising purely from terms in Eq.~\ref{mom3} and Eq.~\ref{mom4}. The suppression is due to an additional factor of $J_{k}(\tb)$ which decays rapidly on long times. This term favors a suppression of occupation at $\pi/2$. Bottom: The momentum distribution at $k=\pi/2$ as a function of time, where the solid and dashed curves represent contributions from same terms as in the top figure.}
\end{figure}

The dynamics of the momentum distribution to second order in $U/J$ is now given by plugging in the expression for the two-body correlation function $g^{pq}_{rs}$ into the equation:
\begin{equation}\label{onebodyfourier}
n_{q}(\tb)-n_{q}(0) =  U \int^{\tb}_{0}dt\int \frac{dp~dr~ds}{(2\pi)^{3}}~\Big(g^{pq}_{rs} - g^{rs}_{pq}\Big)
\end{equation}
One can readily check that the non-interacting expression for $g^{pq}_{rs}$ yields Eq.~\ref{qmtmev}. 

Substituting Eq.~\ref{g2dyn1}, which represents two-particle scattering into the first term on the right hand side of Eq.~\ref{onebodyfourier} one obtains
\begin{widetext}
\begin{equation}\label{mom1}
\int dpdrds \delta_{p+s-r-q}g^{ps}_{rq} = \sum_{k, \delta}\int^{t}_{0}d\tau J^{2}_{\delta}(\tau) \Big[i^{k}J^{2}_{k-\delta}(\tau-t)J_{-k}(t)e^{-i(k q - t \cos{q})} - i^{k-\delta}J^{2}_{k}(-t)J_{\delta-k}(t-\tau)e^{-i((k-\delta)q + (\tau-t)\cos{q})}\Big]
\end{equation}
\end{widetext}

Similarly, the second term in the RHS of Eq.~\ref{onebodyfourier} yields 
\begin{widetext}
\begin{equation}\label{mom2}
\int dpdrds \delta_{p+q-r-s}g^{pq}_{rs} = \sum_{k, \delta}\int^{\tb}_{0}d\tau J^{2}_{\delta}(\tau) \Big[i^{-k+\delta}J_{k-\delta}(\tau-\tb)J^{2}_{-k}(\tb)e^{i((k-\delta)q + (\tau-\tb)\cos{q})} - i^{-k}J_{k}(-\tb)J^{2}_{\delta-k}(\tb-\tau)e^{i(kq - \tb\cos{q})}\Big]
\end{equation}
\end{widetext}

One can readily verify that the RHS of Eq.~\ref{mom2}  is the negative of the complex conjugate of the RHS of Eq.~\ref{mom1}. Hence the final expression is obtained by taking twice the real part of Eq.\ref{mom1}.

A similar analysis for Eq.\ref{g2dyn2}, which arises from three-particle scattering yields for $\int dpdrds \delta_{p+s-r-q}g^{ps}_{rq}$:
\begin{widetext}
\begin{equation}\label{mom3}
\int dpdrds \delta_{p+s-r-q}g^{ps}_{rq} = \sum_{k, \delta}\int^{\tb}_{0}d\tau J^{3}_{\delta}(-\tau) \Big[i^{k}J_{k-\delta}(\tau-\tb)J^{2}_{-k}(\tb)e^{-i(k q - \tb \cos{q})} - i^{k-\delta}J^{3}_{k}(-\tb)e^{-i((k-\delta)q + (\tau-\tb)\cos{q})}\Big]
\end{equation}
\end{widetext}

\begin{widetext}
\begin{equation}\label{mom4}
\int dpdrds \delta_{p+q-r-s}g^{pq}_{rs} =\sum_{k, \delta}\int^{\tb}_{0}d\tau J^{3}_{\delta}(-\tau) \Big[-i^{-k}J_{k-\delta}(\tau-\tb)J^{2}_{-k}(\tb)e^{i(k q - \tb \cos{q})} - i^{-k+\delta}J^{3}_{k}(-\tb)e^{i((k-\delta)q + (\tau-\tb)\cos{q})}\Big]
\end{equation}
\end{widetext}

Once again, the RHS of \ref{mom4} is the negative of the complex conjugate of the RHS of \ref{mom3}

Combining Eqs.~(\ref{mom1}, \ref{mom2}, \ref{mom3} , \ref{mom4}) with the appropriate signs, yields after some manipulation, the final expression in Eq.~\ref{momdistfinal}.

Note that the terms Eqs.~\ref{mom3}, \ref{mom4}, corresponding to three-particle scattering involve an additional factor of $J_{k}$, and are suppressed in magnitude compared to the terms in Eqs.~\ref{mom1}, \ref{mom2}. 

Considered separately these terms have very distinct effects. In Fig.~\ref{momdistquad} we plot the time evolution of the momentum distribution assuming only the second order terms. Note first that both the second order terms yield a momentum distribution that is \textit{symmetric} about $q=\pi/2$, as opposed to the first order term, shown in Fig.~\ref{tofev}. Moreover, while the term representing two-particle scattering tends to \textit{enhance} the peak-like feature at finite momentum, the term representing three-particle scattering tends to \textit{suppress} the occupation at finite momentum.


\begin{thebibliography}{99}

\bibitem{otherbragg} M. Weidem\"uller, A. Hemmerich, A. G\"orlitz, T. Esslinger and T. W. H\"ansch, Phys. Rev. Lett. \textbf{75} 4583 (1995);
G. Raithel, G. Birkl , A. Kastberg, W. D. Phillips and S. L. Rolston, Phys. Rev. Lett. \textbf{78} 630 (1997).

\bibitem{miyakebragg} H. Miyake, G. Siviloglou, G. Puentes, D. E. Pritchard, W. Ketterle and D. M. Weld, Phys. Rev. Lett. \textbf{107} 175302 (2011).

\bibitem{greinerband} M. Greiner, I. Bloch, O. Mandel, T. W. H\"ansch, and T. Esslinger, Phys. Rev. Lett. \textbf{87} 160405 (2001).

\bibitem{demarcoband} D. McKay, M. White and B. DeMarco, Phys. Rev. A \textbf{79}, 063605 (2009).

\bibitem{greinerdyn} W. S. Bakr, J. I. Gillen, A. Peng, M. E. Tai, S. Foelling and M. Greiner, Nature \textbf{462} 74 (2009); W. S. Bakr, A. Peng, M. E. Tai, R. Ma, J. Simon, J. Gillen, S. Foelling, L. Pollet and M. Greiner Science \textbf{329} 547 (2010).

\bibitem{blochquench} J. F. Sherson, C. Weitenberg, M. Endres, M. Cheneau, I. Bloch and S. Kuhr, Nature \textbf{467} 68 (2010). 

\bibitem{chengstruc} C-L. Hung, X. Zhang, L-C. Ha, S-K Tung, N. Gemelke and C. Chin New. J. Phys. \textbf{13} 075019 (2011).

\bibitem{blochnoise} S. F\"olling,  F. Gerbier, A. Widera, O. Mandel, T. Gericke and I. Bloch Nature \textbf{434}, 481 (2005).

\bibitem{altmannoise} E. Altman, E. Demler and M. D. Lukin, Phys. Rev. A \textbf{70} 013603 (2004).

\bibitem{schneidertransport} U. Schneider, L. Hackerm\"uller, J. P. Ronzheimer, S. Will, S. Braun, T. Best, I. Bloch, E. Demler, S. Mandt, D. Rasch and A. Rosch, Nature Physics \textbf{8} 213 (2012); J. P. Ronzheimer, M. Schrieber, S. Braun, S. S. Hodgman, S. Lander, I. P. McCulloch, F. Heidrich-Meisner, I. Bloch and U. Schneider, Phys. Rev. Lett. \textbf{110} 205301 (2013). 

\bibitem{natubogquench} S. S. Natu and E. J. Mueller, Phys. Rev. A \textbf{87} 053607 (2013).

\bibitem{chengsakh} C-L. Hung, V. Gurarie and C. Chin, eprint.arXiv:1209.0011.  

\bibitem{blochlightcone} M. Cheneau, P. Barmettler, D. Poletti, M. Endres, P. Schaub, T. Fukuhara, C. Gross, I. Bloch, C. Kollath and S. Kuhr, Nature, \textbf{481} 484 (2012).

\bibitem{cardy} P. Calabrese and J. Cardy, Phys. Rev. Lett., \textbf{96} 136801 (2006).

\bibitem{liebrobinson} E. H. Lieb and D. W. Robinson, Commun. Math. Phys. \textbf{28} 251 (1972).

 \bibitem{greinercollapse} M. Greiner, O. Mandel, T. W. H\"ansch and I. Bloch, Nature \textbf{419} 51 (2002). 

\bibitem{gerbier} F. Gerbier, A. Widera, S. F\"olling, O. Mandel, T. Gericke and I. Bloch, Phys. Rev. Lett. \textbf{95} 050404 (2005).  

\bibitem{orzelosc} C. Orzel, A. K. Tuchman, M. L. Fenselau, M. Yasuda and M. A. Kasevich, Science \textbf{291} 2386 (2001). 

\bibitem{altman} E. Altman and A. Auerbach, Phys. Rev. Lett. \textbf{89} 250404 (2002).

\bibitem{polkovnikov} A. Polkovnikov, S. Sachdev, S. M. Girvin, Phys. Rev. A \textbf{66} 053607 (2002); 

\bibitem{fischer} U. R. Fischer, R. Schutzhold, M. Uhlmann, Phys. Rev. A \textbf{77} 043615 (2008). 

\bibitem{snoek} M. Snoek, Euro Phys. Lett. \textbf{95} 30006 (2011).

\bibitem{trotzky} S. Trotzky, Y-A. Chen, A. Flesch, I. P. McCulloch, U. Schollw\"ock, J. Eisert, and I. Bloch Nature Physics \textbf{8} 325 (2012).

\bibitem{flesch} A. Flesch, M. Cramer, I.P. McCulloch, U. Schollw\"ock, and J .Eisert  Phys. Rev. A \textbf{78} 033608 (2008).

\bibitem{rigolmtm} M. Rigol and A. Muramatsu, Phys. Rev. Lett. \textbf{93} 230404 (2004).

\bibitem{salvatoremtm} K. Rodriguez, S. R. Manmana, M. Rigol, R. M. Noack and A. Muramatsu, New J. Phys. \textbf{8} 169 (2006). 

\bibitem{demlerexact} P. Barmettler, M. Punk, V. Gritsev, E. Demler and E. Altman, New J. Phys \textbf{12}, 055017 (2010).

\bibitem{sautebd} J. D. Sau, B. Wang and S. Das Sarma, Phys. Rev. A \textbf{85} 013644 (2012).

\bibitem{kollath} A. M. L\"auchli, and C. Kollath, J. Stat. Mech. P05018 (2008).

\bibitem{kollathnoneq} C. Kollath, A. Laeuchli and E. Altman, Phys. Rev. Lett. \textbf{98} 180601 (2007).

\bibitem{kollathferm} P. Barmettler, D. Poletti, M. Cheneau and C. Kollath, Phys. Rev. A \textbf{85} 053625 (2012). 

\bibitem{manmanalightcone} S. Manmana, S. Wessel, R. M. Noack and A.  Muramatsu, Phys. Rev. B \textbf{79} 155104 (2009).

\bibitem{weiss} T. Kinoshita, T. Wenger and D. Weiss, Nature \textbf{440} 900 (2006).

\bibitem{cramer} M. Cramer, C. M. Dawson, J. Eisert and T. J. Osborne, Phys. Rev. Lett. \textbf{100} 030602 (2008).

\bibitem{kehrein} M. Mo\"eckel and S. Kehrein Phys. Rev. Lett. \textbf{100} 175702 (2008).

\bibitem{kollathdecoh} D. Poletti, J-S Bernier, A. Georges and C. Kollath, Phys. Rev. Lett. \textbf{109} 045302 (2012).

\bibitem{greinerbh} M. Greiner, O. Mandel, T. Esslinger, T. W. H\"ansch, and I. Bloch, Nature, \textbf{415} 39 (2002).

\bibitem{fisherbh} M. P. A. Fisher, P. B. Weichman, G. Grinstein and D. S. Fisher, Phys. Rev. B \textbf{40} 546 (1989).






\bibitem{ao} H. T. C Stoof, Phys. Rev. Lett. \textbf{66} 3148 (1991); Phys. Rev. A \textbf{45} 8398 (1992); D. S. Hall, M. R. M Matthews, J. R. Ensher, C. E. Wieman, and E. A. Cornell, Phys. Rev. Lett. \textbf{81} 1539 (1998); P. Ao and S. T. Chui, J. Phys. B \textbf{33} 535 (2000); M. J. Bijlsma, E. Zaremba and H. T. C. Stoof Phys. Rev. A \textbf{62} 063609 (2000).

\bibitem{mukundquench} R. Barnett, A. Polkovnikov and M. Vengalattore, Phys. Rev. A \textbf{84} 023606 (2011). 

\bibitem{spielmancoarsening} S. De, D. L. Campbell, R. M. Price, A. Putra, B. M. Anderson and I. B. Spielman, arXiv eprint: 1211.3127. 

\bibitem{natudemarco} S. S. Natu, D. C. McKay, B. DeMarco and E. J. Mueller, Physical Review A \textbf{85} 061601 (R) (2012).

\end{thebibliography}
\end{document}